\RequirePackage{lineno}
\documentclass[aps,twocolumn,showpacs,showkeys,amsmath,amssymb,superscriptaddress,groupedaddress]{revtex4}  
\usepackage{graphicx}  
\usepackage{dcolumn}   
\usepackage{bm}        
\usepackage{amsmath,amssymb}   
\usepackage{multirow}
\usepackage{indentfirst}
\usepackage{longtable}
\usepackage{epsfig}
\usepackage{graphicx}
\usepackage{times}
\usepackage{latexsym}
\usepackage{mathrsfs}
\usepackage{color}
\begin{document}

\vspace*{-2cm}
\hspace*{-3cm}
Submitted to 'Chinese Physics C'

\title{A Precise Calculation of Delayed Coincidence Selection Efficiency and Accidental Coincidence Rate}
\author{Jingyi Yu}
\author{Zhe Wang\footnote{Correspondence: wangzhe-hep@mail.tsinghua.edu.cn}}
\author{Shaomin Chen}
\affiliation{Center for High Energy Physics,\\ Department of Engineering Physics,\\
             Tsinghua University, Beijing, China}
\date{\today}
\begin{abstract}
    A model is proposed to address issues on the precise background
    evaluation due to the complex data structure defined by the delayed
    coincidence method, which is widely used in reactor
    electron-antineutrino oscillation experiments.
    In this model, the effects from the muon veto, uncorrelated random
    background, coincident signal and  background are all studied with the
    analytical solutions, simplifying the estimation of the systematic
    uncertainties of signal efficiency and accidental background rate
    determined by the unstable single rate. The result of calculation is
    validated numerically with a number of simulation studies and is also applied and validated in the recent Daya Bay hydrogen-capture based oscillation measurement.

\end{abstract}


\maketitle

\section{Introduction}
    The delayed coincidence method is used very broadly in nuclear and high energy physics experiments. The existence of a delayed signal greatly relaxes the critical requirement on the random background level. Recently three reactor electron-antineutrino experiments Double CHOOZ~\cite{DC1st, DC2, DCnH}, RENO~\cite{RENO1st} and Daya Bay~\cite{DYB1st, DYB2, DYB3, DYBnH} adopted this established technique to precisely measure the neutrino mixing angle $\theta_{13}$~\cite{pontecorvo, mns}. Electron-antineutrinos from reactors were distinguished by detecting the coincidence of the prompt positron and delayed neutron capture signals of inverse-beta-decay IBD interactions, $\bar{\nu}_{e}+p \rightarrow e^{+}+n$.
    The expected high precision of the $\sin^22\theta_{13}$ measurements required a better understanding of the acceptance of delayed-coincidence signals and their background contamination, especially for the situation when the backgrounds are much higher if using the neutron capture signals on hydrogen (nH)~\cite{DCnH, DYBnH}, instead of the capture on gadolinium (nGd).

    This article describes a complete mathematical model with analytical solutions for these neutrino experiments, which addresses realistic muon veto, high accidental background and varying single rate.
    They were not completely discussed by the formula with the neutron-like signal rate, Eq. 5 of~\cite{DC2} or the off-window method~\cite{DCnH}.
    It improves the understanding of delayed-coincidence signals and backgrounds, and was particularly applied to one recent Daya Bay study~\cite{DYBnH}. It is also useful to other on-going and future reactor neutrino experiments to study the neutrino mass hierarchy~\cite{hierarchy, hierarchy2, hierarchy3, hierarchy4}.

    These three experiments have a very similar design. The antineutrino detectors all have a three-layer structure, where the central region is filled with gadolinium-doped liquid scintillator GdLS, the middle region with pure liquid scintillator LS, and the outmost layer with mineral oil. The IBD signals are categorized according to neutron capture nuclei. In the GdLS region, $\sim$84\% of neutrons capture on Gd and release several gammas with a total energy of $\sim$8~MeV, while the rest are dominantly captured on hydrogen and release a 2.2~MeV gamma. The average capture time is $\sim$29~$\mu s$. In the LS region, most neutrons capture on hydrogen and release a 2.2~MeV gamma, but the average capture time is much longer and is $\sim$200~$\mu s$. Two types of dedicated analyses are in progress. The nGd analyses have the best precision while the nH analyses make a non-trivial contribution.

    In these analyses, the full data-taking time is chopped into fragments by cosmic muons. Take the Daya Bay nH analysis as an example for explanation. The total muon rate at the near site was $\sim$200~Hz. Different veto windows were applied after each muon, \emph{i.e.}\ 1~s, 800~$\mu s$ or 400~$\mu s$ depending on the energy deposition of the muon track and its distance to the detector. A longer veto window is needed when the energy deposition of a muon is larger, because more long-lived spallation backgrounds may be found immediately after it~\cite{Kamland}. The total vetoed time took $\sim$20\% at the near site. It is useful to understand the distribution for the time fragment durations between these muon vetoes.

    The background situation is complicated and particularly very high for the nH analyses. For the Daya Bay case, four major backgrounds were identified, which included accidental coincidence background, spallation $^9$Li/$^8$He and fast neutron backgrounds induced by cosmic-ray muons, and $^{241}$Am-$^{13}$C calibration source background. The goal of the precision of $\sin^22\theta_{13}$ is less than one percent, however, the accidental background statistics is comparable to the IBD statistics for both the Double Chooz and Daya Bay far-site detectors in the nH analyses. Can these backgrounds be described by a complete mathematical model along the time axis? Can the systematic uncertainty of the accidental background be controlled to sub-percent?

    Besides these concerns, other issues from the random background rate also deserve attention in the nH analyses. The random background was decreasing slowly in the Daya Bay case, since it is caused by residual radioactivity, which decreased after the antineutrino detectors were immersed in water. After muons, the random rate also increase quickly, because muons may introduce a lot of spallation backgrounds.

\section{Data Model and Simplification}
\label{sec:II}

\subsection{Data model}
\label{subsec:DM}
    A model was built up to address issues on the precise background
    evaluation due to the complex data structure. In the following we use $e$ and $n$ to represent the positron and neutron from an IBD reaction respectively, $s$ a random background (or single) and $\mu$ for a muon. The features of each type of event are described below.

    1) The event of central importance is the delayed-coincidence IBD event, including a prompt signal $e$ and a delayed signal $n$. The event rate of IBD's is assumed to be $R_{\rm IBD}$. The time between the prompt and delay signals can be complicated. For the three neutrino experiments mentioned above, the neutron capture cross section varies with its momentum, so that an exponential distribution is a poor approximation for the neutron capture time. An abstract form $P_{delay}(t)$ is used to represent the delayed signals' time distribution.

    All $^9$Li/$^8$He, fast neutron and $^{241}$Am-$^{13}$C backgrounds have a correlated delayed signal caused by a real neutron. For the $^9$Li/$^8$He background, neutrons are emitted from the beta-delayed neutron decay of $^9$Li or $^8$He. the fast neutron background's prompt signal is the recoil of the neutron. Similarly, the $^{241}$Am-$^{13}$C background's prompt signal is also from the recoil of the neutron from the source. In terms of delayed-coincidence selection, they are not distinguishable from the IBD events, so that they are all categorized as coincident events and presented as IBDs in later discussion. Details on how to identify these coincident backgrounds with energy information, etc.\ can be found elsewhere~\cite{DC1st, RENO1st, DYB1st, DC2, DCnH, DYB2, DYB3, DYBnH, Kamland} and is beyond the scope of this paper.

    2) Random signals, or singles, include decays from residual radioactive nuclei in the detector and from the environment, or other non-correlated detector noise. Random signals occur with a uniform distribution,
   {\it i.e.} the time interval between two random signals follows an exponential distribution with an average value of $1/R_s$, where $R_s$ is the single rate.

    3) With the inclusion of cosmic ray muons, all elements are covered with the model.

\begin{figure*}
      \centering
      \includegraphics[width=14cm]{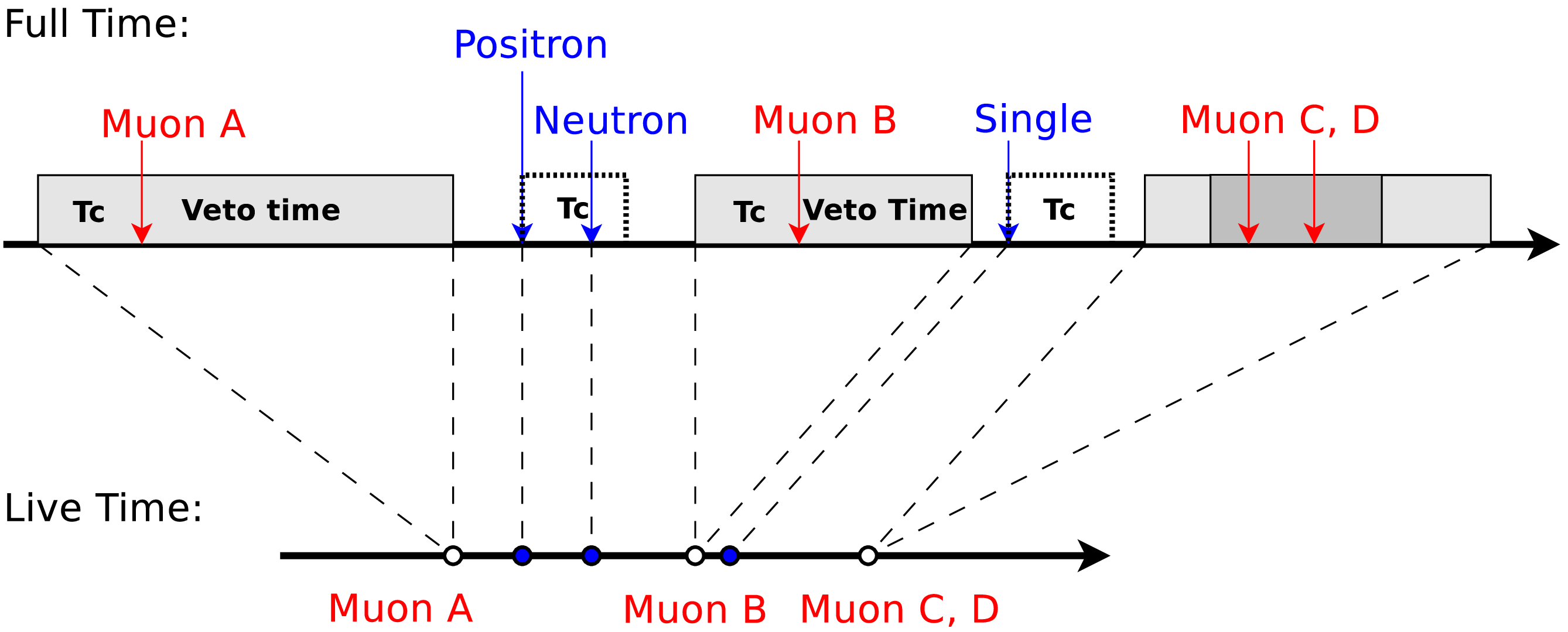}
      \caption{(Color online) On the top is a full-time axis, and from the left to right are muon A, a pair of delayed-coincidence events (positron and neutron), muon B, random single, and muon C and D. A fixed-length coincidence window $T_c$ is opened for each possible candidate, i.e. the positron and the single. The total dead time for prompt signals is $T_c$ plus the muon veto window (the shaded area). In this example, muon A has a longer veto window than B, and muon C and D's veto windows overlap. On the bottom is a live-time axis where the dead time of each muon is contracted as a single point and all other events are unchanged. More details can be found in Section~\ref{sec:II}.}
      \label{fig:LiveTime}
    \end{figure*}

    These different types of signals in the full data-taking time axis are shown in Fig.~\ref{fig:LiveTime}, as well as one type of delayed coincidence searching method (multiplicity two selection). First come two muon events, A and B, on the full-time axis. Muon A is supposed to be closer to the sensitive region of the detector than muon B, and so, a longer veto window is applied to muon A. One pair of delayed coincidence signals (a positron and a neutron) occurs between muon A and B. A fixed-length coincidence window $T_{c}$ is opened after the positron which is a possible prompt signal candidate, since it is not vetoed by any muons or occupied by any other coincidence-searching windows. Window $T_{c}$ is usually comparable to the average arrival time of the delayed signal, for example 400~$\mu s$ for nH analyses. For a delayed signal to not fall into a muon veto window, the time between a prompt signal and the next muon event cannot be smaller than $T_{c}$. The total dead time introduced by a muon is the veto time plus $T_{c}$ for prompt candidates. After muon B, a single event occurs, since no delayed signal is found within $T_{c}$. Finally two muons, C and D, are very close to each other in time, so that their veto windows overlap. In this example all coincident signals and singles are well separated. The situations with overlaps are discussed in the following sections.

\subsection{Simplification}
\label{subsec:Simp}
    The first attempt of simplification is to project all signals from the full-time axis to the live-time axis. As illustrated in Fig.~\ref{fig:LiveTime}, a muon with its dead time is contracted as a point on the live-time axis and counted as one net muon, like muon A and B. Muons C and D have overlapping dead times, so they are contracted together as one point on the live-time axis and counted as only one net muon. All other signals are simply moved to the live-time axis. Since every muon's dead time is already removed from the live-time axis, the full length of the live-time axis is the total time of the prompt signal searching.

    The important assumption making this analytic calculation possible is to postulate that the net muons are uniformly distributed on the live-time axis and the time interval follows an exponential distribution with an average value of $1/R_{\mu}$ where $R_{\mu}$ is the net muon rate on the live-time axis.

    In the calculation below, we always have $R_{\rm IBD} \ll R_s$. For the three reactor neutrino experiments mentioned, $R_s$ is about four orders of magnitude higher than $R_{\rm IBD}$.

\section{Calculation Method}
\label{sec:III}
\subsection{Delayed coincidence events and other combinations}
    Besides the necessary predictions of the event rates of delayed-coincidence signals and the accidental background, other types of combinations may also be of interest. They are grouped according their multiplicity.

    One-fold coincident events: $s$, $e$, $n$.

    Two-fold coincident events: $ss$, $se$, $sn$, $en$, $es$, $ns$.

    Three-fold coincident events: $sss$, $sse$, $ses$, $sns$, $ssn$, $sen$, $ens$, $esn$, $ess$, $nss$.

\subsection{Two-step calculation}
    The calculation of these event rates is divided into two steps: a) determine the probability of a type of signal to start a coincidence searching window; b) determine the probability that there is a second or third signal in the searching window for two-fold or three-fold coincidences or that there is no other signal for one-fold events. In the following sections, the starting probabilities of a single background ($P_{s-start}$), positron ($P_{e-start}$) and neutron ($P_{n-start}$) will be calculated first, followed by the rates of all kinds of combinations.

\subsection{Starting probability}
    On the live-time axis, each signal except a muon can start a coincident searching window, as long as it is not in the previous coincidence searching window. Note that on the live-time axis, $R_{\rm IBD}$ and $R_s$ are exactly the same as on the full-time axis. A single event may start a searching window in different situations. In the formulas below, $t_{\mu}$ is the time to its previous net muon event.

\begin{figure}
  \centering
  \includegraphics[width=\columnwidth]{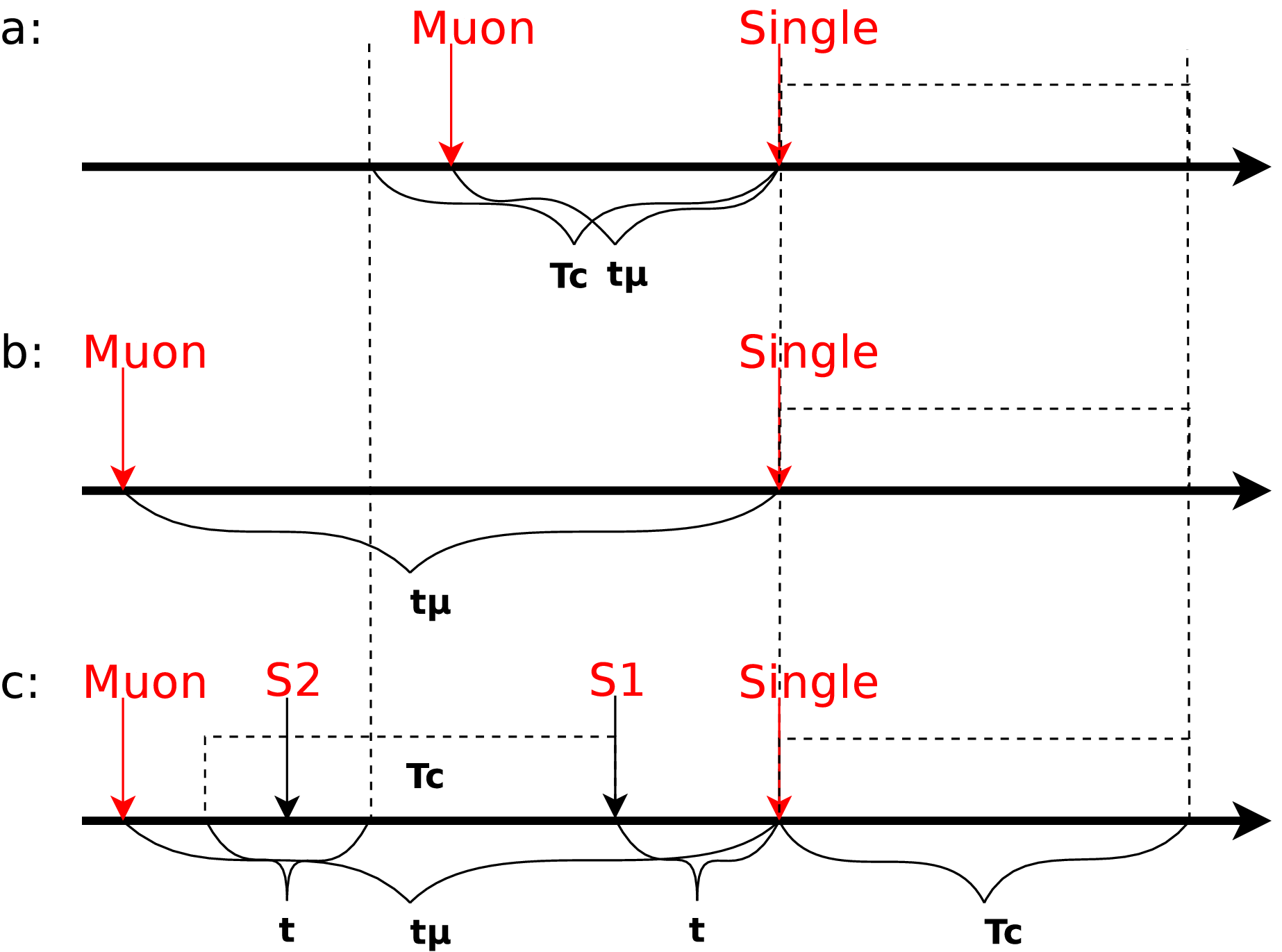}
  \caption{(Color online) Three situations of how a target `Single' event starts a searching window. For case a, that the time to previous muon $t_{\mu}$ is less than the coincidence window $T_c$, and no event occur within $t_{\mu}$ ensures that the target single is not within any muon veto or previous coincidence windows. In case b, $t_{\mu}$ is longer than $T_c$, and a searching window can start as long as it is not occupied by a previous coincidence window, {\it i.e.}\ nothing occurs within $T_c$. Case c shows an extension based on b, the prior single `S1' is very close, but is within its previous searching window opened by `S2'. See the text in Sec.~\ref{sec:III} for details. }
  \label{fig:Pstart}
\end{figure}

    Case a) As shown in Fig.~\ref{fig:Pstart} a, when $t_{\mu}<T_{c}$, if there is no other signal between the single event and the muon, a searching window will be started by the single event. The probability of this situation is
    \begin{align}
    P_{a} &= \int_0^{T_{c}} P(0|R_{s}t_{\mu}) \cdot P(t_{\mu}) \cdot d t_{\mu} \nonumber\\
          &= \int_0^{T_{c}} \frac{(R_s t_{\mu})^k}{k!} e^{-R_s t_{\mu}} \vert_{k=0} \cdot R_{\mu} e^{-R_{\mu}t_{\mu}} \cdot dt_{\mu} \nonumber\\
          &= \frac{R_{\mu}}{R_{s}+R_{\mu}}[1-e^{-(R_{s}+R_{\mu})T_{c}}],
    \end{align}
    where $P(0|R_{s}t_{\mu})$ is the probability that there is no other single event in between, which is calculated as a Poisson distribution with a mean of $R_s t_{\mu}$ and count $k=0$, and the second term $P(t_{\mu})$ gives the probability of finding a muon at $t_{\mu}$ before the target single event being considered, which is calculated according to an exponential distribution with a rate of $R_{\mu}$.

    Case b) When $t_{\mu}\geqslant T_{c}$, if there is no other signal within $T_{c}$ before the target single event, a searching window will be started as the panel b of Fig.~\ref{fig:Pstart} shows. The starting probability of this situation is
    \begin{align}
    P_{b} &= P(0|R_{s}T_{c}) \cdot \int_{T_{c}}^{\infty} P(t_{\mu}) \cdot dt_{\mu} \nonumber\\
          &= \frac{(R_s T_{c})^k}{k!} e^{-R_s T_{c}} \vert_{k=0} \cdot \int_{T_{c}}^{\infty} R_{\mu} e^{-R_{\mu}t_{\mu}} \cdot dt_{\mu} \nonumber\\
          &= e^{-(R_{s}+R_{\mu})T_{c}},
    \end{align}
    where the first term corresponds to no signals in $T_{c}$ and the second term is to ensure $t_{\mu} \geqslant T_{c}$.

    Case c) is an extension of b). When $t_{\mu}>T_{c}$, there is another scenario that the target single can start a searching window as depicted in panel c of Fig.~\ref{fig:Pstart}. There is a random signal, $s1$, before the target single event, however it occurs within a previous searching window started by $s2$. The probability of this situation is
    \begin{align}
    P_{c} =& \int_{0}^{T_{c}}dtR_{s}e^{-R_{s}t}\cdot [1-P(0|R_{s}t)] \cdot \int_{T_{c}+t}^{\infty}P(t_{\mu})\cdot dt_{\mu} \nonumber\\
          =& \frac{R_s}{R_s+R_{\mu}}e^{-R_{\mu}T_c}[1-e^{-(R_s+R_{\mu})T_c}] \nonumber\\
           & -\frac{R_s}{2R_s+R_{\mu}}e^{-R_{\mu}T_c}[1-e^{-(2R_s+R_{\mu})T_c}],
    \end{align}
    where the first integral gives the probability to find the first random background signal $s1$ at time $t$ before the target single event, the second term calculates the probability of having at least one random background $s2$ in the early $t$ window, that is $T_c$ away from the target single event, and the last integral just gives the probability of finding a muon at some time larger than $T_c+t$.

    Finally, the starting probability is
    \begin{equation}
    P_{s-start}=P_{a}+P_{b}+P_{c}.
    \end{equation}
    There should be higher order corrections after $P_{c}$, but, with the example parameters used in the simulation presented later, they are estimated to be five orders of magnitude smaller than these leading terms.

    The starting rate is
    \begin{equation}
        R_{s-start}=R_{s}\cdot P_{s-start}.
    \end{equation}


    For the prompt signals of IBD events, the starting probability is different from $P_{s-start}$ by only a positron detection efficiency $\varepsilon_{e}$:
    \begin{equation}
    P_{e-start}=P_{s-start}\cdot\varepsilon_{e}.
    \end{equation}
    Then the rate of searching windows started with positrons is
    \begin{equation}
    R_{e-start}=R_{\rm IBD}\cdot P_{s-start} \cdot \varepsilon_{e}.
    \end{equation}

    The situation of a neutron starting a searching window is a bit complicated, because by nature it is always possible that a prompt positron signal is ahead of it. It only happens when the positron is failed to be detected. The method to calculate $P_{n-start}$ is the same as above and is not shown. The event rate with neutrons as a start is expressed as
    \begin{equation}
    R_{n-start}=R_{\rm IBD} \cdot P_{n-start}.
    \end{equation}

\subsection{Construct an event}
    After the starting probability and the starting rate of a type of signal are known, the rate of accidental background, detectable IBD pairs and other cases can be calculated.

    The accidental background rate $R(ss)$ is just the rate of one single to start a searching window multiplied by the probability of a second single event appearing in the same window
    \begin{equation}
    R(ss)=R_{s-start}\cdot P(1|R_{s}T_{c}),
    \end{equation}
    where $P(1|R_{s}T_{c})$ is the Poisson probability of one count with a mean of $R_{s}T_{c}$. IBD events are not explicitly required to be excluded from this because $R_{\rm IBD} \ll R_s$.

    The detectable IBD event rate $R(en)$ can be obtained in a similar way:
    \begin{align}
    R(en) =& R_{e-start} \cdot \varepsilon_{n|e} \cdot \int_{0}^{T_{c}}P_{delay}(t)dt\cdot P(0|R_{s}T_{c})\nonumber\\
          =& R_{\rm IBD}\cdot P_{s-start}\cdot \varepsilon_{e} \cdot \varepsilon_{n|e} \cdot  \nonumber\\
           & \int_{0}^{T_{c}}P_{delay}(t)dt \cdot P(0|R_{s}T_{c}),
    \end{align}
    where $\varepsilon_{n|e}$ is the efficiency of neutron detection after a positron is detected, the integral is the time cut efficiency, and the random background is explicitly required to be outside this searching window in the last term. The complete IBD event detection efficiency can be expressed as:
    \begin{equation}
    \varepsilon_{\rm IBD} = \frac{R(en)}{R_{\rm IBD}}.
    \end{equation}
    It can be found that $\varepsilon_{e}$, $\varepsilon_{n|e}$, and time cut related efficiency $P_{delay}(t)$ integral can all be factored out and studied separately. The details of $P_{delay}(t)$ will not affect the selection.
    The result also applies to other non-IBD correlated backgrounds. The two-fold selection efficiencies of them are the same to IBD events.
    The results of other combinations are similar and not shown.

\section{Verification with Monte Carlo}
\label{sec:IV}

    Since the muon veto cut, etc. were quite complicated, high statistics ($1\times10^{10}$ events) Monte Carlo simulation studies were done to verify the predictions. Three types of events were produced: IBD, single, and muon. They were generated on the full-time axis according to three uniform distributions. Several sets of parameters were tried according to the real case as in~\cite{DYBnH}.
    The muon rate on the full-time axis was set to the highest muon rate of 200~Hz as in the Daya Bay near site, since any muon rate lower than this would have a less significant effect if the model or simplification failed. Usually a veto time of 400~$\mu s$ was applied for each muon, but 0.05\% of them can be shower muons, for which a one-second long veto was applied. The shower muon fraction 0.05\% is close to the real case, and the shower muon dead time takes up $\sim$10\% of the full time. The shower fraction had been tested up to 1\%, and the total dead time almost covered the entire full-time axis. The neutron capture time distribution $P_{delay}(t)$ was represented with a simple exponential distribution with rate $\lambda$. The detection efficiencies of the prompt and delayed signals were also included. One set of parameters used is summarized in Table~\ref{tab:param}.

    \begin{table*}[]
    \begin{tabular}{c c c c c c c c c c c }
      \hline
       $R_s$ & Muon rate & Veto time & Shower fraction & Shower veto & $R_{\rm IBD}$ &$\epsilon_{e}$&$\epsilon_{n|e}$&$T_c$ & $1/\lambda$ \\\hline
       50 Hz & 200 Hz & 400 $\mu$s & 0.05\% & 1 s & 0.1 Hz & 1 & 0.8 & 400 $\mu$s & 200 $\mu$s\\\hline
    \end{tabular}
    \caption{Monte Carlo simulation parameters. From left to right, they are single rate, muon rate,
    muon veto window, shower muon fraction, shower veto window, IBD rate, prompt signal detection efficiency,
    conditional detection efficiency of neutron, coincidence window and average neutron capture time.}
    \label{tab:param}
    \end{table*}

    \begin{table*}[]
    \begin{tabular}{l c c c c c c c c}
      \hline
     Rate [Hz] &	$R(s)$&$R(ss)$    &$R(se)$      &$R(sn)$     &$R(sss)$    &$R(ens)$   &$R(sse)$    &$R(sns)$  \\
               &        &         &           &          &          &+$R(esn)$  &+$R(ses)$   &+$R(ssn)$ \\\hline
      Mea.     &48.0833 &0.96165  &0.0010460  &0.0001108 &0.009626  &0.001342 &2.10E-5   &2.27E-6\\
      Sta. Err.& 0.0015 &0.00022  &7.2E-6     &2.3E-6    &2.2E-5    &1.1E-5   &1.4E-6    &4.7E-7\\
      Pred.    &48.0856 &0.96171  &0.0010499  &0.0001093 &0.009617  &0.001330 &2.10E-5   &2.19E-6\\
      Diff.    & -1.5   &-0.27    &-0.54      &0.65      &0.41      &1.1      &0         &0.17   \\\hline
      Rate [Hz]&$R(e)$    &$R(en)$    &$R(es)$      &$R(sen)$    &$R(ess)$   &$R(n)$      &$R(ns)$     &$R(nss)$\\\hline
      Mea.     &0.029657&0.066447 &0.0005991  &0.0008715 &6.91E-6  &0.012719  &0.0002616 &3.11E-6\\
      Sta. Err.&0.000038&0.000057 &5.4E-6     &6.6E-6    &5.8E-7   &2.5E-5    &3.6E-6    & 3.9E-7\\
      Pred.    &0.029647&0.066525 &0.0005930  &0.0008735 &5.93E-6  &0.012770  &0.0002554 &2.55E-6\\
      Diff.    &0.26    &-1.4     &1.1        &-0.30     &1.7      &-2.0      &1.7       &1.4    \\
      \hline
    \end{tabular}
    \caption{Measured event rates of simulation sample and their predictions for all one-, two-, and three-fold cases. For each type of combination, Mea. gives the measured result, Sta. Err. is its statistical error, Pred. is the predicted value, and Diff. is (Mea.-Pred.)/Err.
    The situation $ens$ is not distinguished with $esn$ in the analysis, as well as $sse$ with $ses$, and $sns$ with $ssn$.}
    \label{tab:comp}
    \end{table*}

    The measurements with the simulated sample and the predictions are listed in Table~\ref{tab:comp}, where the discrepancies are all within a 3-$\sigma$ range.
    It was found that the net muon we defined still follows Poisson statistics on the live-time axis and the net muon rate
    on the live-time axis and the real muon rate on the full-time axis are the same within the statistical uncertainty.

\section{Application}
\label{sec:V}
    In the recent Daya Bay nH analysis~\cite{DYBnH}, the method was implemented with real data. The muon rate or net muon rate can be measured with data, and both were observed to be stable throughout the period. A precise single rate can also be extracted. In the Daya Bay case, there are several kinds of real coincidence events, IBD, and $^9$Li/$^8$He, fast neutron and $^{241}$Am-$^{13}$C backgrounds, as pointed out earlier, and their total rates were known to be four orders of magnitude lower than the singles rate. So an upper limit was calculated as a trigger rate of everything excluding muons. And a safe lower limit was calculated by further rejecting all correlated-like events if two triggers were too close in time and in distance. (Although the correlated background rates of $^{214}$Bi-$^{214}$Po-$^{210}$Pb and $^{212}$Bi-$^{212}$Po-$^{208}$Pb were high, they were rejected before the multiplicity selection by a 1.5 MeV energy cut for every trigger.) Taking the average of the upper and lower limits, a precise estimation of $R_s$ was obtained with a systematic of 0.18\%, 0.16\% and 0.05\% for the three sites of the Daya Bay experiment, respectively, which reflected the real coincidence event fractions at each site. After an estimation of all correlated events is obtained, an iteration can further improve the accuracy.
    For Daya Bay, $R_s$ was observed unstable in two aspects. $R_s$ decreased slowly at a rate ($<0.36$\%/day), because the singles were originally from residual radioactivity in the detector and after it was sealed, the total number could decrease. Another effect was the instant increase after muons, since muons may introduce some spallation products. With a veto window of a few hundreds of micro seconds, some of them can still survive.
    With these values as input, the IBD efficiency and accidental backgrounds rate, etc.\ can all be calculated. The uncertainties in IBD detection efficiency and accidental background can be directly derived according the systematic uncertainty in the $R_s$ and its variance as a function of real time and as a function of the time to previous muon.

    Validation with data is also possible. As in~\cite{DYBnH}, the distance distributions of all two-fold events were studied, as well as the time distributions. It was known that only accidental coincidences can have a large separation distance ($>2$~m) or a long coincidence time ($>1.5$~ms). On the other hand, a single sample can be selected and randomly combined to predict the spectrum of the accidental background. With the normalization constant provided by the formulas above, the predicted accidental spectrum can be compared with the data as in Fig. 2 of~\cite{DYBnH}. Given the high statistics of the Daya Bay data, the systematics of accidental backgrounds were validated to $<0.2$\%, which is sufficient for the expected $\sin^22\theta_{13}$ measurement precision.

\section{Conclusion}
\label{sec:VI}
    A complete mathematical model was developed for the signal and background distributions on the full-time axis for delayed-coincidence experiments, for example, recent reactor neutrino experiments. It was then projected onto the live-time axis. An analytic calculation was done by assuming the net muons are uniformly distributed on the live-time axis, and the real correlated signals have a much smaller rate than the single rate.
    The intrinsic relative uncertainties for all combinations' rates are at the $10^{-5}$ level. The predictions were verified with high statistics simulation studies with realistic parameters from the Daya Bay experiment.
    The model was also applied to the Daya Bay nH analysis and validated to high precision with data.
    With analytical expressions, it is convenient to consider the systematic uncertainties of the IBD detection efficiency and the accidental background rate when the single rate is unstable.
    In the future large reactor antineutrino experiments aiming to resolve the neutrino mass hierarchy~\cite{hierarchy, hierarchy2, hierarchy3, hierarchy4}, to achieve a longer attenuation length in liquid scintillator, the nH method is preferred over the nGd method, $i.e.$\ to not use Gd-load liquid scintillator. The method presented here is directly applicable for these studies.

\section{Acknowledgement}
    The work is supported in part by the Ministry of Science and Technology of China (Grant No. 2013CB834302), the National Natural Science Foundation of China (Grant No. 11235006 and 11475093), the Tsinghua University Initiative Scientific Research Program (Grant No. 2012Z02161), and the Key Laboratory of Particle \& Radiation Imaging (Tsinghua University), Ministry of Education.



\end{document}